\documentstyle{europhys}

\def\And{{\rm and\ }}

\newif\ifboo \boofalse

\def\Review#1{\boofalse{\it #1},}
\def\Name#1{{\sc #1},}
\def\Vol#1{\ifboo Vol. {\bf #1}\else{\bf #1}\fi}
\def\Year#1{\ifboo #1\else(#1)\fi}

\def\Page#1{\ifboo {\rm p. #1}\else{\rm #1}\fi}

\begin{document}
\euro{}{}{}{}
\Date{}
\shorttitle{Yang G. and Maki K. Impurity Scattering in f-wave ...
   }

\title{Impurity Scattering in f-wave Superconductor UPt$_3$}
\author{Guolin Yang\inst{1} \And Kazumi Maki\inst{1}}
\institute{
     \inst{1} Department of Physics and Astronomy,\\
     University of Southern California\\
     Los Angeles,CA90089-0484, USA\\
     }
\rec{}{}
\pacs{
\Pacs{74}{20-h}{Theory}
\Pacs{74}{25Bt}{Thermodynamic properties}
\Pacs{74}{25Fy}{Transport properties}
  }    
\maketitle

\begin{abstract}
We study theoretically the effect of impurity scattering in f-wave (or
$E_{2u}$) superconductors. The quasi-particle density of states of f-wave
superconductor is very similar to the one for d-wave superconductor as
in hole-doped high T$_c$ cuprates. Also in spite of anisotropy in
$\vec{\Delta}(\hat{k})$, both the normalized superfluid density and
the thermal conductivity is completely isotropic.
\end{abstract}

\section{Introduction}
After a long controversy, f-wave (or $E_{2u}$) superconductivity in
UPt$_3$ is established in 1996\cite{1}. First the thermal
conductivities with the heat current parallel to the c axis and within
the basal plane decrease linearly in T the temperature at low
temperature\cite{2}. This is consistent with $E_{2u}$ but inconsistent
with the then prevailing model $E_{1g}$\cite{3,4}. Almost at the same time
$^{195}$Pt Knight shift measurement in UPt$_3$ finds the spin triplet
pair\cite{5}. Later they find also among three phases A, B
and C in UPt$_3$, only B phase is nonunitary\cite{5}. This is again
consistent with $E_{2u}$ but not with $E_{1g}$. 

So we can write down the superconducting order parameter 
\begin{equation}
\vec{\Delta}(\hat{k})=\alpha \Delta\hat{d}\hat{k_3}(\hat{k1}\pm
i\hat{k_2})^2
\end{equation}
and $\alpha=\frac{3\sqrt{3}}{2}$, $\hat{d}\parallel\vec{c}$ and
$\vec{k}$ is the quasi-particle momentum.

Very recently we have shown that f-wave superconductivity describes
quite well the observed upper critical field in UPt$_3$\cite{6,7}.

The object of this letter is to study the impurity effect in f-wave
superconductivity. It is well known that impurity produces profound
effect in unconventional superconductors\cite{8,9}. Also for the analysis of
transport properties the impurity scattering is crucial. Assuming that
the impurity scattering is in the unitarity limit, we analyze the
thermodynamic and transport properties. In particular we find both the
normalized superfluid density and the thermal conductivity are
completely isotropic.

The present result describes very well the temperature dependence of
the ultrasonic attenuation coefficient of UPt$_3$\cite{10}. Also the
deviation from the universal limit in the low temperature thermal
conductivity\cite{11} in f-wave superconductor is very similar but 
 somewhat larger
than in d-wave superconductor\cite{8}, which may be consistent with
the recent measurement of electron-irradiated UPt$_3$\cite{12}.

\section{Formulation}
Following the standard approach the effect of impurity scattering is
incorporated by replacing the frequency $\omega$ in the quasi-particle
Green function in the Nambu space by the renormalized one
$\tilde\omega$.

\begin{equation}
G^{-1}(\omega,\vec{p})=\tilde{\omega}-\xi \rho_3-\alpha\Delta\rho_1
k_3(k_1\pm i k_2\rho_3)^2 \sigma_1
\end{equation}
and
\begin{equation}
\tilde\omega=\omega+i
\Gamma<\frac{\tilde\omega}{\sqrt{\tilde\omega^2-\Delta^2 f^2}}>^{-1}
\end{equation}
where $\rho_i$ are the Pauli matrices in the Nambu space,
$\Gamma=n_i/\pi N_0$ the scattering rate, $f=\alpha
\sin^2\theta\cos\theta$ and $<\cdots>$ means the average on the Fermi
surface.

Then the Gap equation is given by
\begin{equation}
\lambda^{-1}=2\pi T\frac{1}{<|f|^2>}\sum^\prime_n<\frac{|f|^2}
{\sqrt{\tilde\omega^2_n+\Delta^2|f|^2}}>
\end{equation}
here $<f^2>=\frac{18}{35}$, $\lambda$ is the dimensionless coupling
constant, and $\tilde{\omega_n}$ is the renormalized Matsubara
frequency. When $\Delta\rightarrow 0$, we obtain the universal
Abrikosov-Gor'kov formula\cite{13}
 \begin{equation}
-\ln(\frac{T_c}{T_{c0}})=\psi(\frac{1}{2}+\frac{\Gamma}{2 \pi
 \Gamma_c})-\psi(\frac{1}{2})
\end{equation}
where $T_c(T_{c0})$ is the superconducting transition temperature in
the presence (absence) of impurities. Also we have $T_c=0$when
$\Gamma=\Gamma_c=\frac{\pi}{2\gamma}T_{c0}=0.8819 T_{c0}$. For
$T=0K$, Eq(4) reduced to
\begin{eqnarray}
\displaystyle -\ln(\frac{\Delta(\Gamma,0)}{\Delta_{00}})=&\displaystyle \frac{1}{<f^2>}\left\{<f^2\ln(\frac{
C_0+\sqrt{C_0^2+f^2}}{f})>-C_0<\frac{f^2}{\sqrt{C_0^2+f^2}}>+\right.\nonumber\\
&\displaystyle \left. +\zeta\int_{C_0}^\infty
 du<\frac{f^2}{(u^2+f^2)^{3/2}}><\frac{1}{\sqrt{u^2+f^2}}>^{-1}\right\}
\end{eqnarray}
where $\zeta=\frac{\Gamma}{\Delta}$ and $iC_0=u(\equiv
\frac{\tilde{\omega}}{\Delta})|_{\omega=0}$ is given by
\begin{equation}
C_0^2=\zeta<\frac{1}{\sqrt{C_0^2+f^2}}>^{-1} \Rightarrow \sqrt{3}\zeta\big
[\ln(\frac{1}{C_0})+const\big ]^{-1}
\end{equation}
where the last expression is the limiting value for $\zeta\rightarrow
0$.

Also the residual density of states is given by 
\begin{equation}
\frac{N(0)}{N_0}=C_0<\frac{1}{\sqrt{C_0^2+f^2}}>=\frac{\zeta}{C_0}=
\frac{\Gamma}{\Delta C_0}
\end{equation}
We show in Fig.1 $T_c/T_{c0}, \Delta(\Gamma,0)/\Delta_{00}$, and
$N(0)/N_0$ versus $\Gamma/\Gamma_c$. This figure is strikingly similar
to the one in d-wave superconductor\cite{8}.

The quasi-particle density of states is given by
\begin{equation}
\frac{N(E)}{N_0}=Re<\frac{\tilde{\omega}}{\sqrt{\tilde{\omega}^2-\Delta^2f^2}}>
\end{equation}
where we put $E=\omega$ in Eq(3). This is shown for a few
$\zeta=\Gamma/\Delta$ versus $E/\Delta$ in Fig2a). We display in Fig.2b)
the corresponding one for d-wave superconductor\cite{14}. They are
strikingly similar except for large value of $\zeta$. For larger $\zeta$,
$N(E)/N_0$ for f-wave superconductor approaches  the normal state
result much faster.

Also Eq(4) is solved numerically and we show
$\Delta(\Gamma,T)/\Delta_{00}$ for a few $g=\Gamma/\Gamma_c$ versus
$T/T_{c0}$ in Fig.3.

\section{Thermodynamics}
It is convenient to start  with entropy given by
\begin{eqnarray}
S=-4\int^\infty_0 dE N(E)\big[f\ln f+(1-f)\ln(1-f)\big]~~~~~~~\nonumber\\
=~~ 4\int^\infty_0 dE N(E)\big[\beta E(1+e^{\beta
  E})^{-1}+\ln(1+e^{-\beta E})\big]
\end{eqnarray}
Then we obtain
\begin{equation}
\frac{H_c^2(T)}{8\pi}=\int^{T_c}_T dT(S_n(T)-S(T))
\end{equation}
and $S_n(T)=\frac{2\pi^2}{3}N_0T$ the entropy in the normal state and
$H_c(T)$ is the thermodynamic critical field. In Fig.4 we show
$D(\frac{T}{T_c})$ versus $(T/T_c)^2$, where
\begin{equation}
D(\frac{T}{T_c})=\frac{H_c(T)}{H_c(0)}-(1-(\frac{T}{T_c}))
\end{equation}
for several $g=\Gamma/\Gamma_c$.

Also the specific heat is given by $C_s=T\frac{dS}{dT}$. In Fig.5 we
show $C_s/\gamma_n T$ where $\gamma_n=\frac{2\pi^2}{3}N_0$. The
specific heat thus obtained is quite consistent with the
observation\cite{15}. We are concerned here with the phase
$T_{c-}$. Perhaps the more details of the theory can be tested within
the system with controlled impurity concentration\cite{12}.

Finally the normalized  superfluid density is isotropic and given 
by 
\begin{equation}
\rho_s(\Gamma,T)=2\pi T\sum_{n=0}^\infty<\frac{\Delta^2
  f^2}{(\tilde\omega_n^2 + \Delta^2 f^2)^{3/2}}>
\end{equation}
This follows from the relation
\begin{equation}
3\int^1_0z^2F(f)=\frac{3}{2}\int^1_0(1-z^2)F(f)=\int^1_0dz F(f)
\end{equation}
where F is an arbitrary function of $f=\frac{3\sqrt{3}}{2}z(1-z^2)$.

First at $T=0K$, Eq(13) reduces to
\begin{equation}
\rho_s(\Gamma,0)=
1-\frac{N(0)}{N_0}+\zeta\int^\infty_{C_0}du\Big(\frac{\displaystyle
  <\frac{f^2}{(u^2+f^2)^{3/2}}>}{\displaystyle
  <\frac{u}{\sqrt{u^2+f^2}}>}\Big)^2 
\end{equation}
This is shown in Fig.6 versus $\Gamma/\Gamma_c$ together with the one
in d-wave superconductor. They are almost indistinguishable one from
the other. 

Finally $\rho_s(\Gamma,T)$ for several $g$ versus $\frac{T}{T_c}$ is
shown in Fig.7. In the pure limit $\rho_s(\Gamma, T)$ decrease
linearly with $T$ as in d-wave superconductor\cite{8,16}.

\section{Transport Properties}
Following\cite{17} the ultrasonic attenuation coefficient for the
transverse wave with $\vec{q}\parallel\vec{b}$ and
$\vec{e}\parallel\vec{c}$ and $\vec{q}\parallel\vec{b}$ and
$\vec{e}\parallel\vec{a}$ are given by
\begin{equation}
\frac{\alpha_{sc}}{\alpha_N}=\frac{15\zeta}{2}\int^\infty_0\frac{dE}{2T}
\mbox{sech}^2(\frac{E}{2T})\int^1_0dz
z^2(1-z^2)\frac{h(u,f)}{Im\sqrt{u^2-f^2}}
\end{equation}
and
\begin{equation}
\frac{\alpha_{sa}}{\alpha_N}=\frac{15\zeta}{8}\int^\infty_0\frac{dE}{2T}
\mbox{sech}^2(\frac{E}{2T})\int^1_0dz
(1-z^2)^2\frac{h(u,f)}{Im\sqrt{u^2-f^2}}
\end{equation}
respectively, where
\begin{equation}
h(u,f)=\frac{1}{2}(1+\frac{|u|^2-f^2}{|u^2-f^2|})
\end{equation}
and $u=\tilde\omega/\Delta$ and $\omega=E$ in Eq(3).

Eq(16) and Eq(17) are evaluated for several g and shown in Fig.8 and 9
respectively. They are quite consistent with the experimental
data\cite{10}.

Finally the thermal conductivity is given by
\begin{equation}
\frac{\kappa_s(T)}{\kappa_n(T)}=\frac{3\Gamma}{2\pi^2\Delta}\int^\infty_0\frac{dE}{2T^3}
E^2\mbox{sech}^2(\frac{E}{2T})\frac{h(u,f)}{Im\sqrt{u^2-f^2}}
\end{equation}
which is shown in Fig.10. Perhaps of more interest is
\begin{equation}
\frac{\kappa}{\kappa_0}=\lim_{T\rightarrow
  0}\frac{\kappa_s(T)}{\kappa_{s0}(T)}=\frac{\sqrt{3}\Delta_{00}}
{\Delta(\Gamma,0)}<\frac{C_0^2}{(C_0^2+f^2)^{3/2}}>
\end{equation}  
which is shown versus g in Fig.11. This describes the deviation from
the universality $(\frac{\kappa}{\kappa(\Gamma,0)}=1$ for
$g\rightarrow 0)$. We note the deviation is a little greater than the one for
d-wave superconductors\cite{8}. 

\section{Summary} In summary we have studied the effect of impurity
scattering in f-wave superconductors. We found f-wave superconductor
behaves in many respects very similar to d-wave superconductor.

Also we predict the impurity dependence of the thermodynamic
properties and transport properties of f-wave superconductors, which
should be readily accessible to experiments. Finally the specific heat
data from another triplet superconductor URu$_2$Si$_2$\cite{18}
appears very consistent with f-wave superconductor.
 
\section{Acknowledgment}
We thank J. P. Brison, and J. Flouquet, T. Ishiguro and Y. Maeno for
providing us the experimental data prior to publication. Of of us (KM) 
thanks Max-Planck Institut f\"ur Physik Komplexer Systeme at Dresden for
hospitality and continued support.

\newpage          
 \begin{center}
 Figure caption
\end{center}

\noindent
Figure 1. $T/T_{c0},\Delta(\Gamma,0)/\Delta_{00}$ and
$N(0)/N_0$ versus $\Gamma/\Gamma_c$.

\vspace{0.5cm}
\noindent
Figure 2. The quasi-particle density states $N(E)/N_0$ versus
$E/\Delta$ for several $\Gamma/\Delta$. a) for f-wave, b) for d-wave
superconductor respectively.

\vspace{0.5cm}
\noindent
Figure 3. $\Delta(\Gamma,T)/\Delta_{00}$ versus $T/T_c$ for several
$g=\Gamma/\Gamma_c$.

\vspace{0.5cm}
\noindent
Figure 4. $D(T/T_c)$ versus $(T/T_c)^2$. In the pure limit
$|D(T/T_c)|$ is much larger than the one for s-wave superconductor. But
$|D(T/T_c)|$ decrease monotonically with increasing $g=\Gamma/\Gamma_c$. 

\vspace{0.5cm}
\noindent
Figure 5. $C_s/\gamma_N T$ versus $T/T_{c0}$.

\vspace{0.5cm}
\noindent
Figure 6.$\rho_s(\Gamma,0)$ versus $\Gamma/\Gamma_c$. This behavior is
very similar to the one in d-wave superconductor. 

\vspace{0.5cm}
\noindent
Figure 7. $\rho_s(\Gamma,T)$ versus $T/T_{c0}$ for several $g$.

\vspace{0.5cm}
\noindent
Figure 8. $\alpha_s/\alpha_N$ versus $T/T_{c0}$ for several $g$ for
$\vec{q}\parallel \vec{b}$ and $\vec{e}\parallel\vec{c}$.

\vspace{0.5cm}
\noindent
Figure 9. The same as Fig.8 for $\vec{q}\parallel \vec{b}$ and
$\vec{e}\parallel\vec{a}$.

\vspace{0.5cm}
\noindent
Figure 10.$\kappa_s(T)/\kappa_n(T)$ versus $T/T_{c0}$ for several
$g$. Here $\kappa_n(T)=\frac{\pi^2 n}{3 m \Gamma}T$, and n is the electron
density.

\vspace{0.5cm}
\noindent
Figure 11.$\kappa/\kappa_0=\lim_{T\rightarrow
  0}\kappa_s(T)/\kappa_{s0}(T)$ versus $\Gamma/\Gamma_c$. Here
  $\kappa_{s0}(T)=\lim_{\Gamma\rightarrow 0}\kappa_s(T)$,
  $\kappa/\kappa_0 > 1$ implies the deviation for the universality.
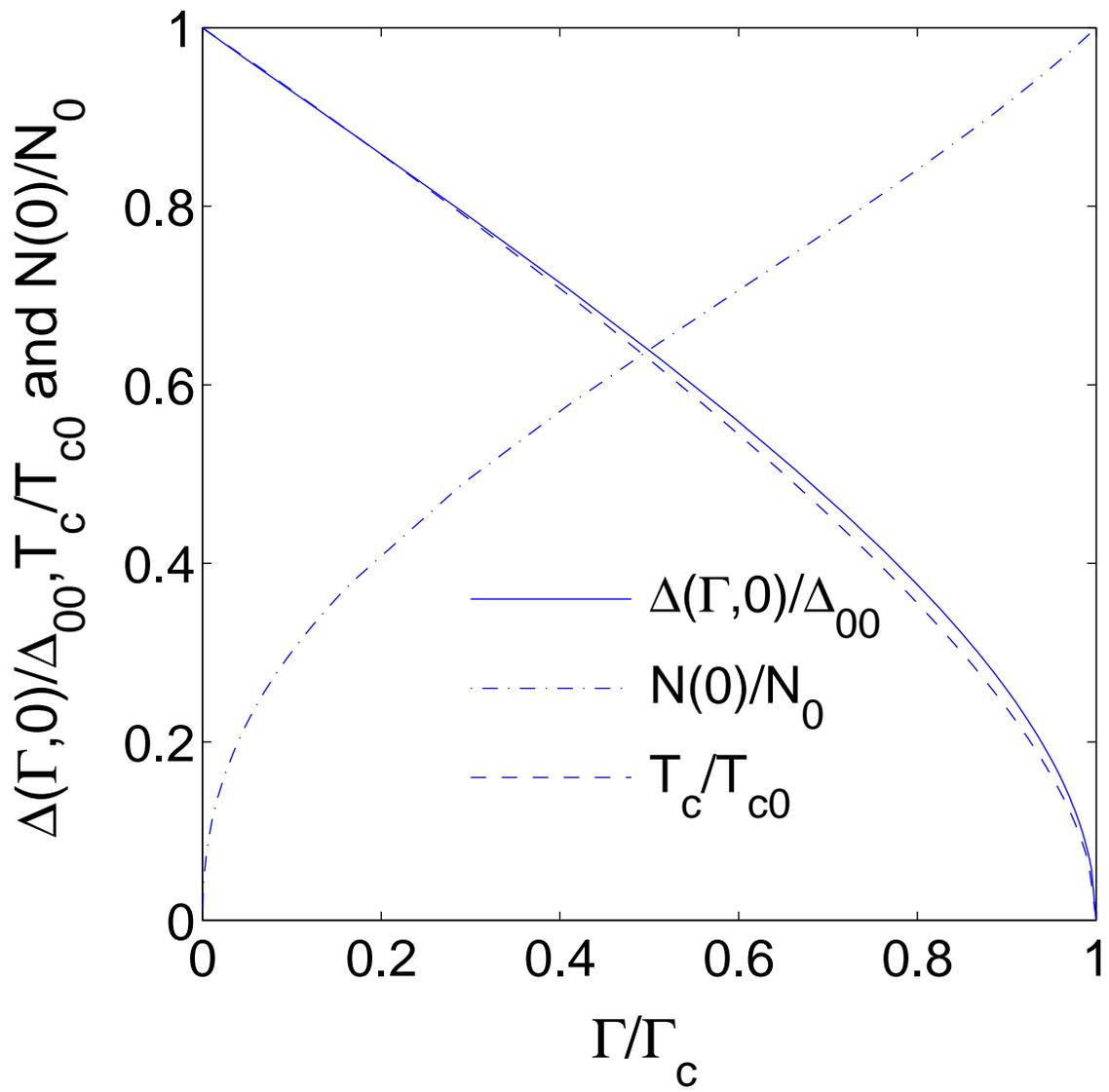
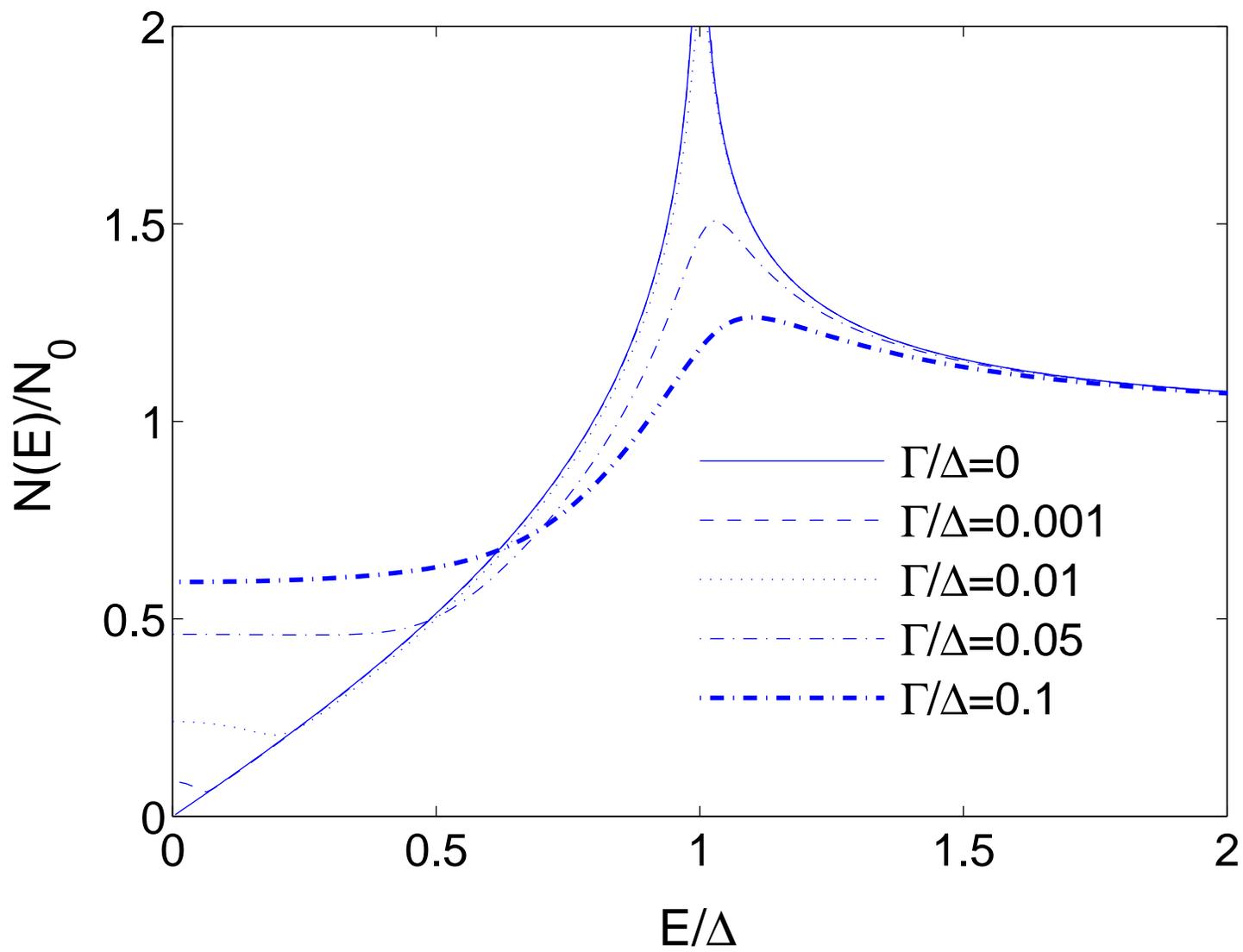
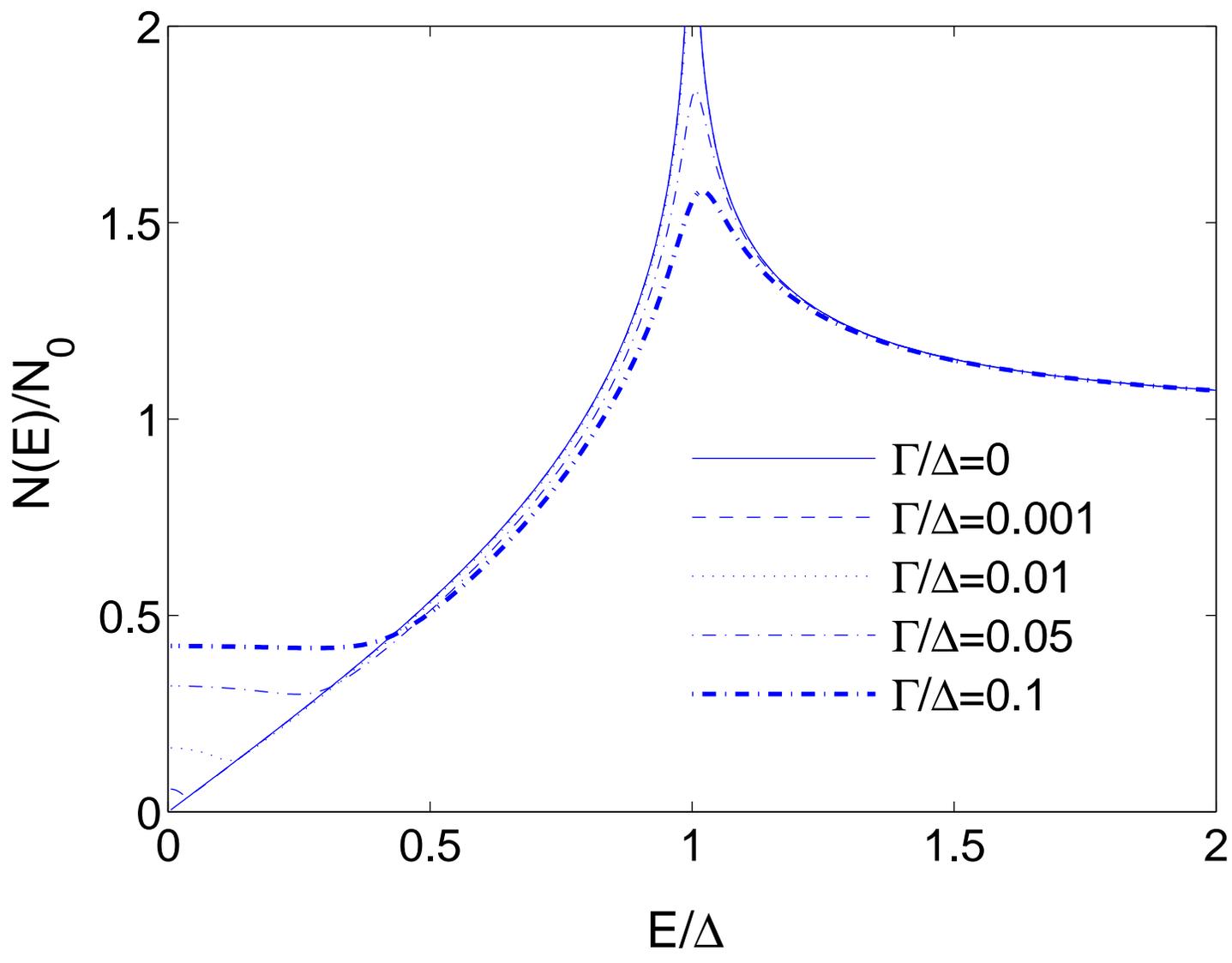
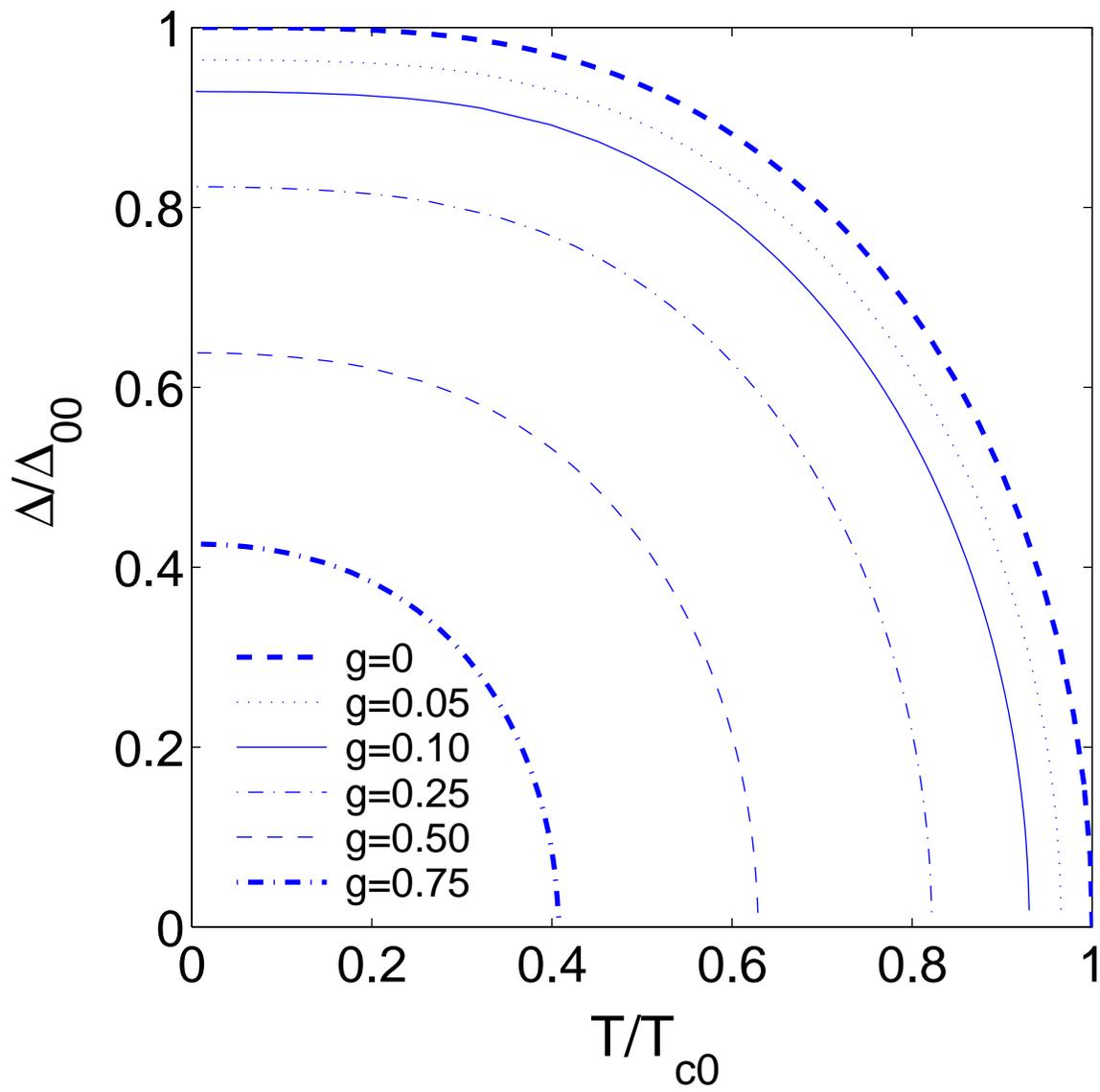
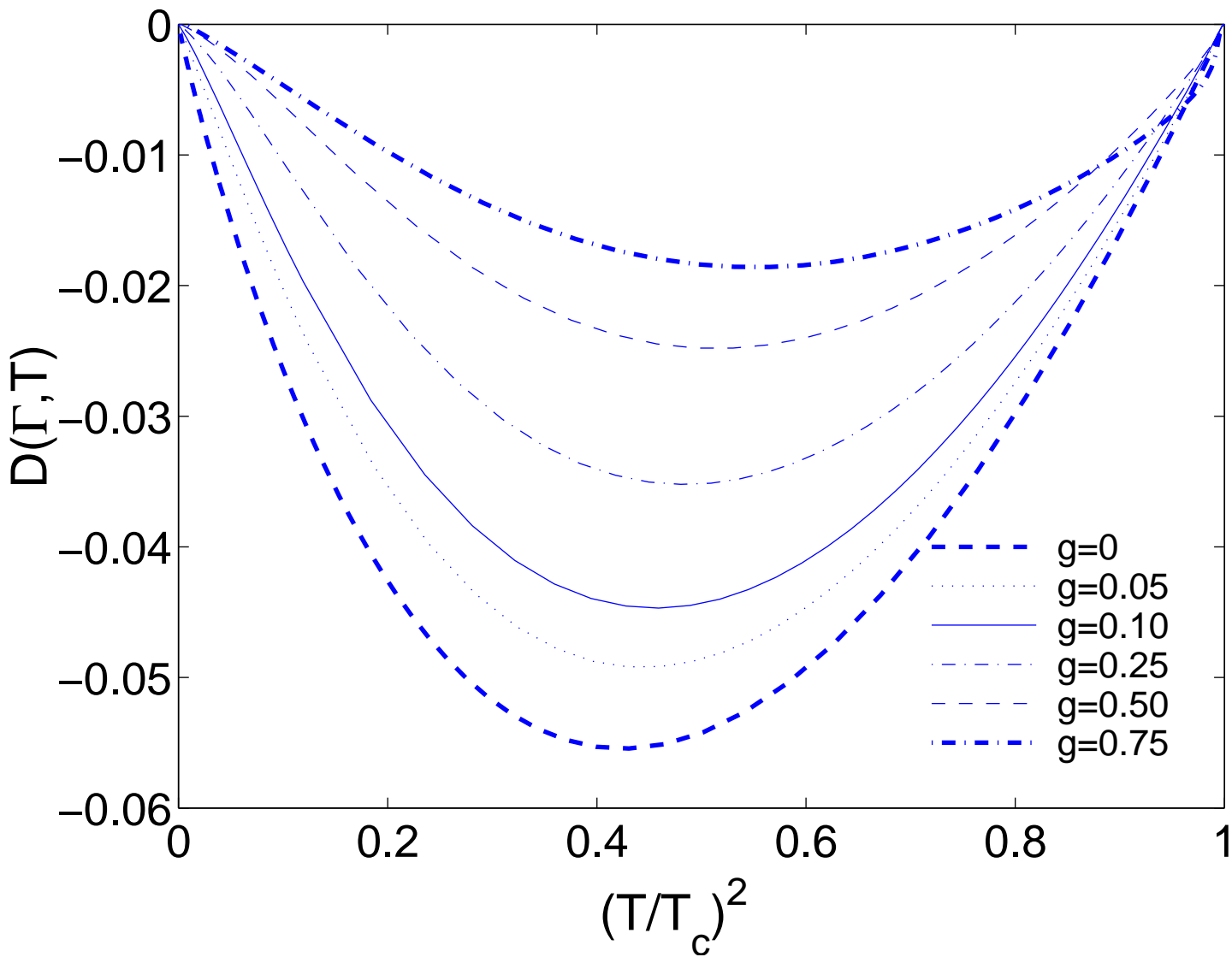
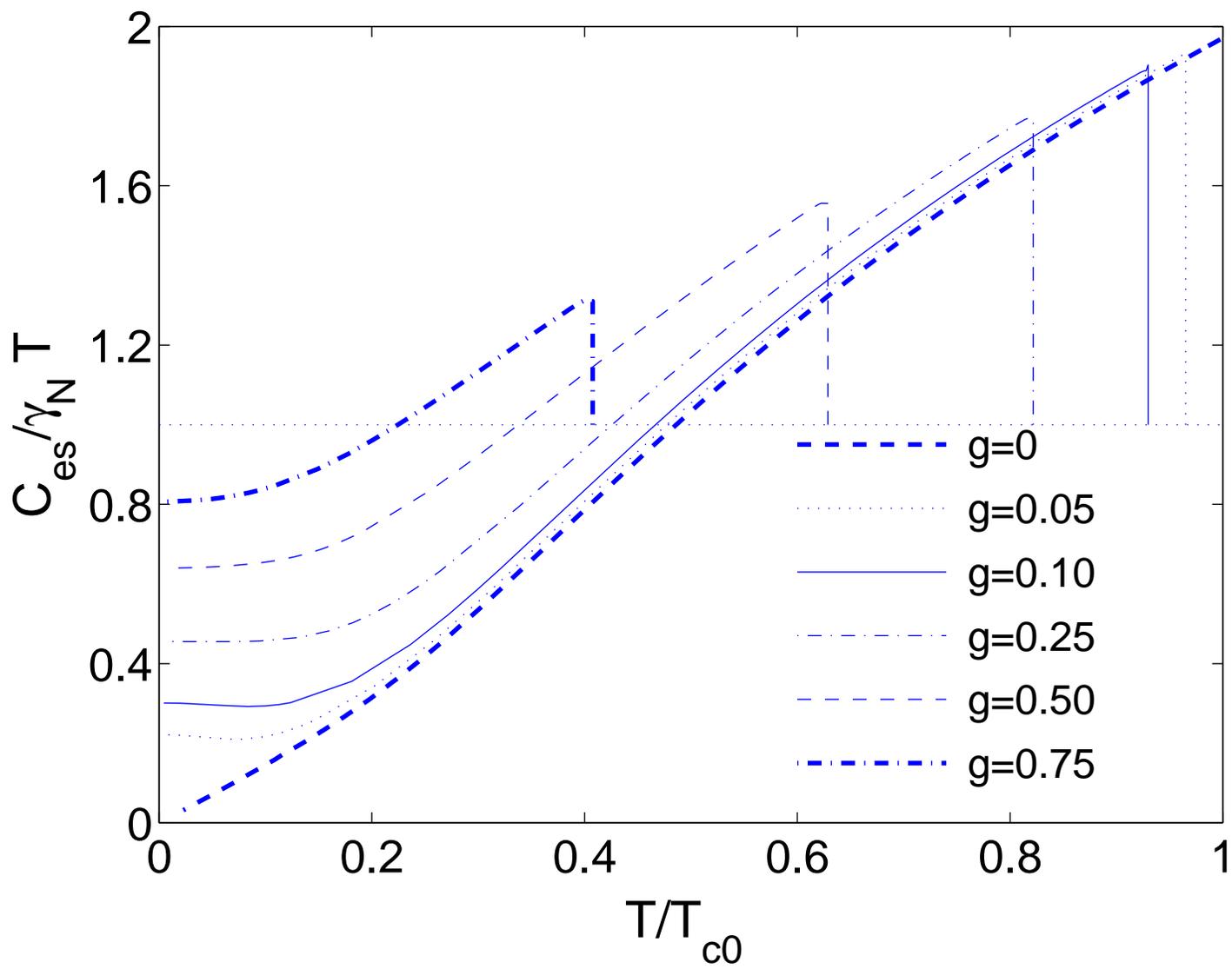
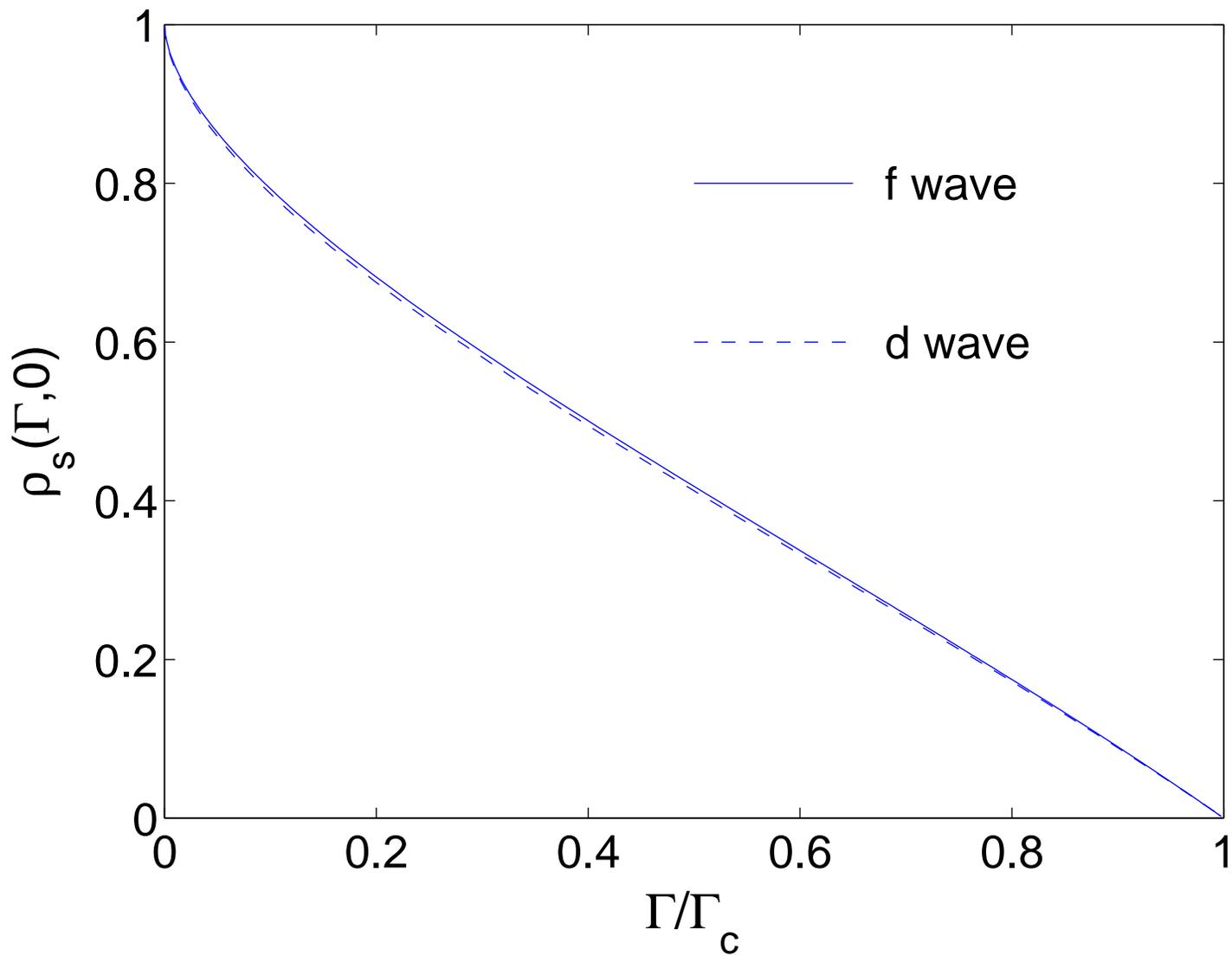
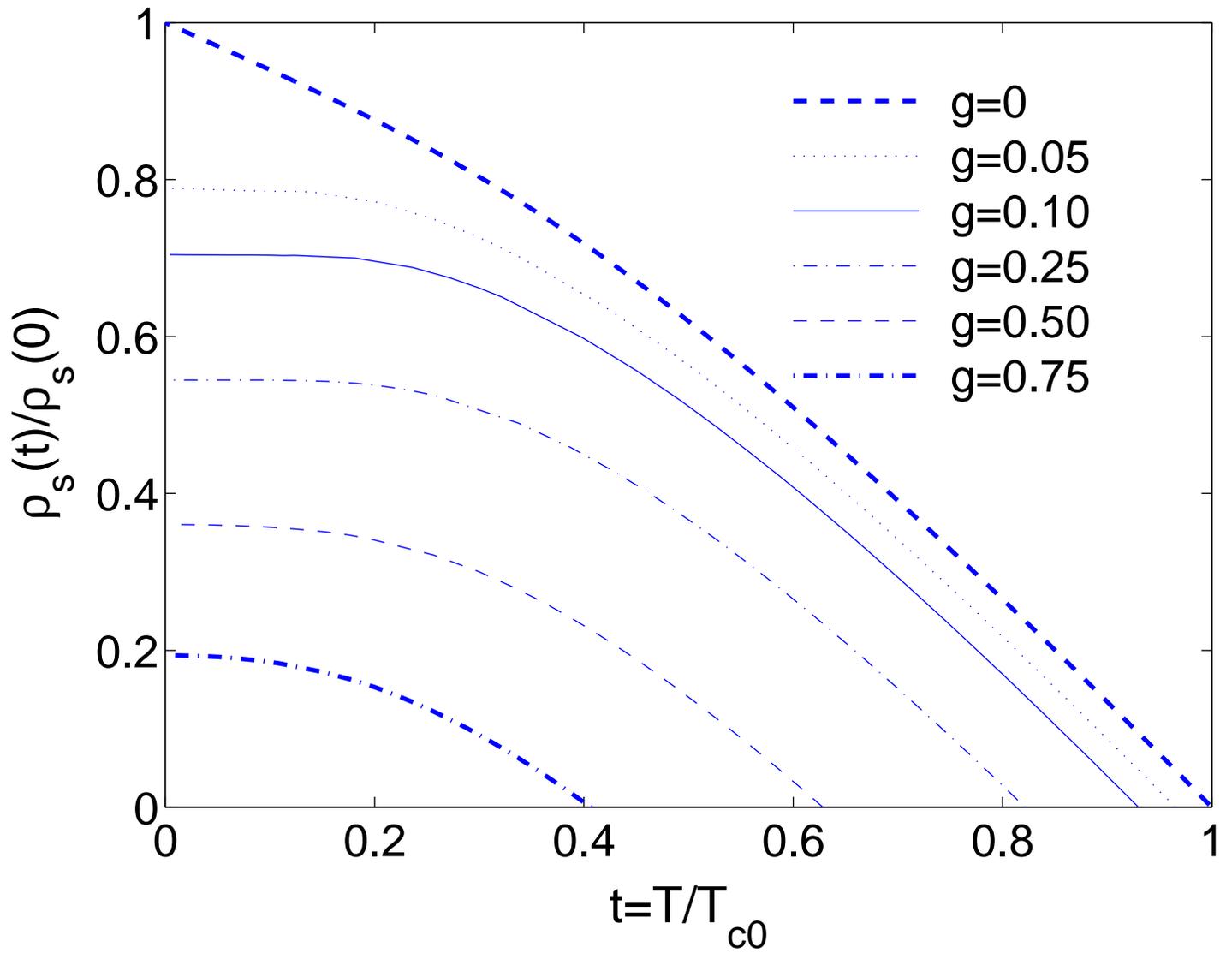
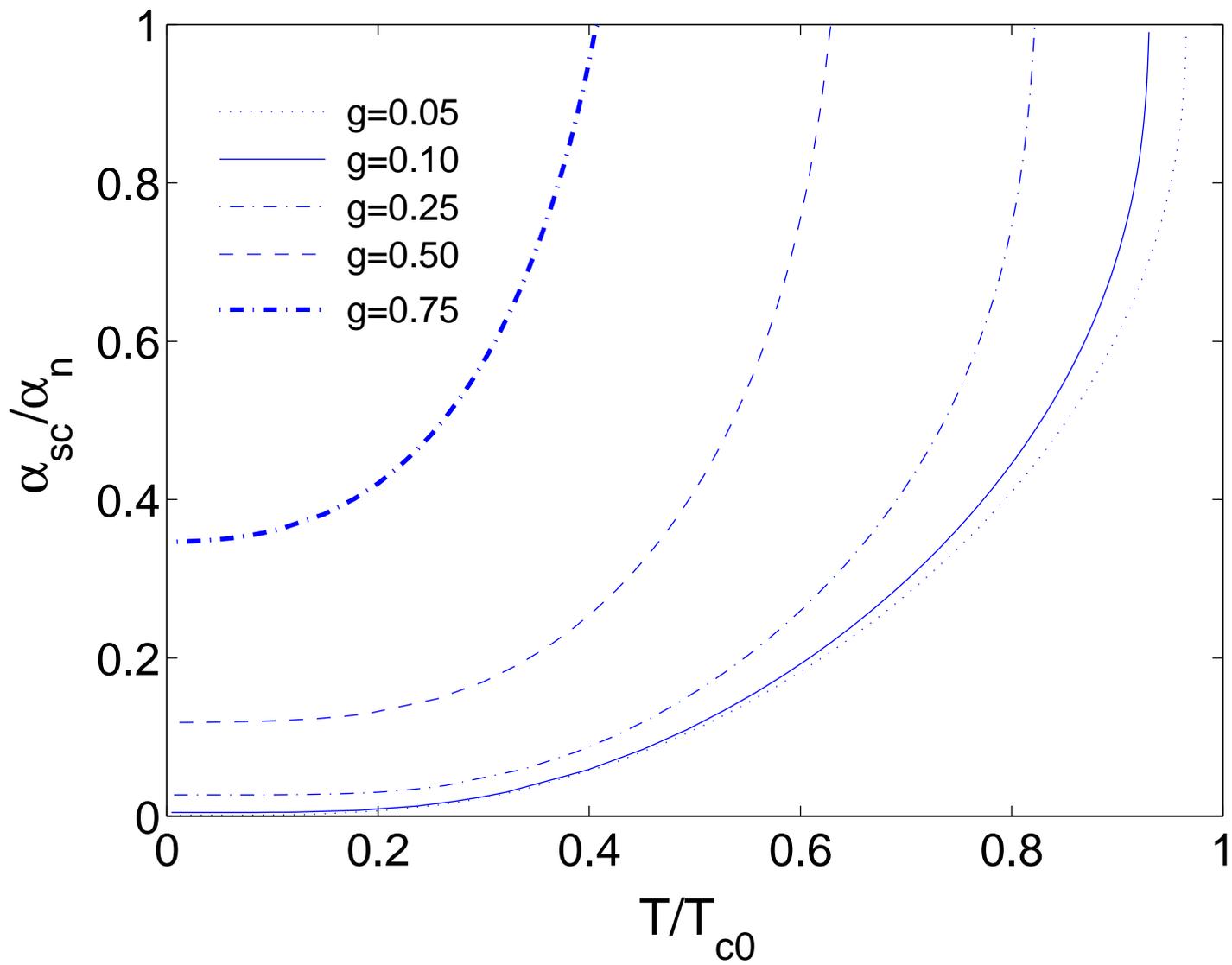
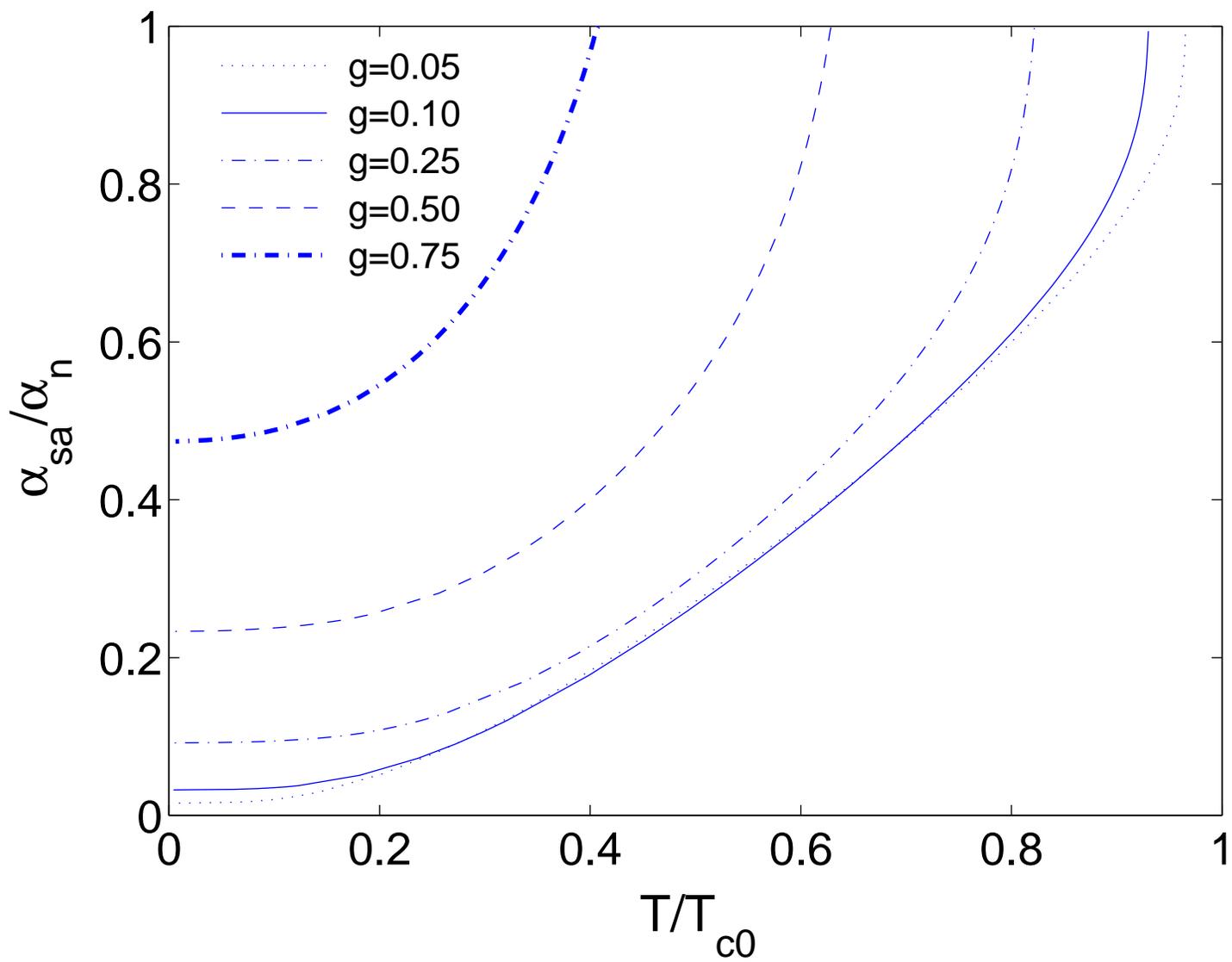
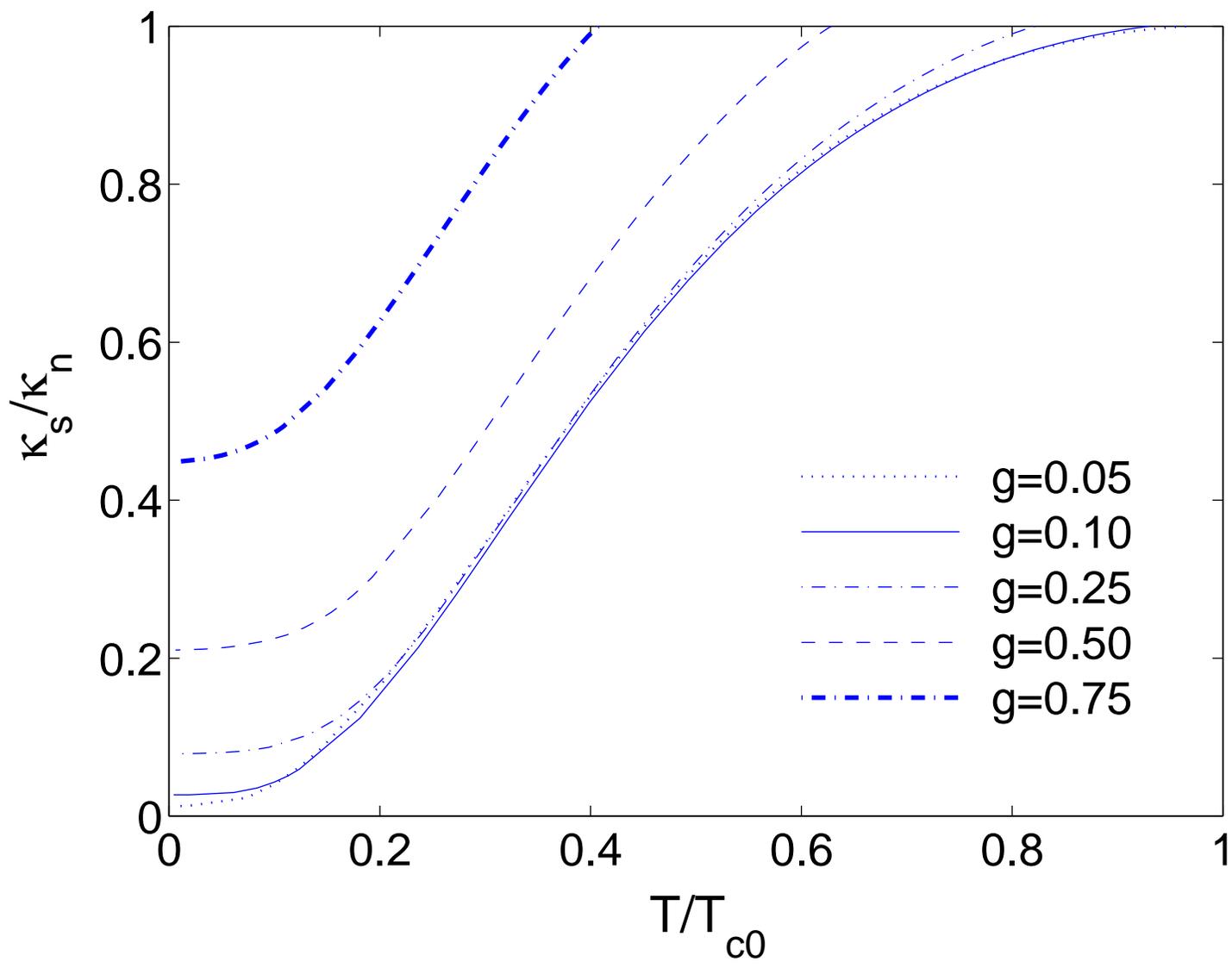
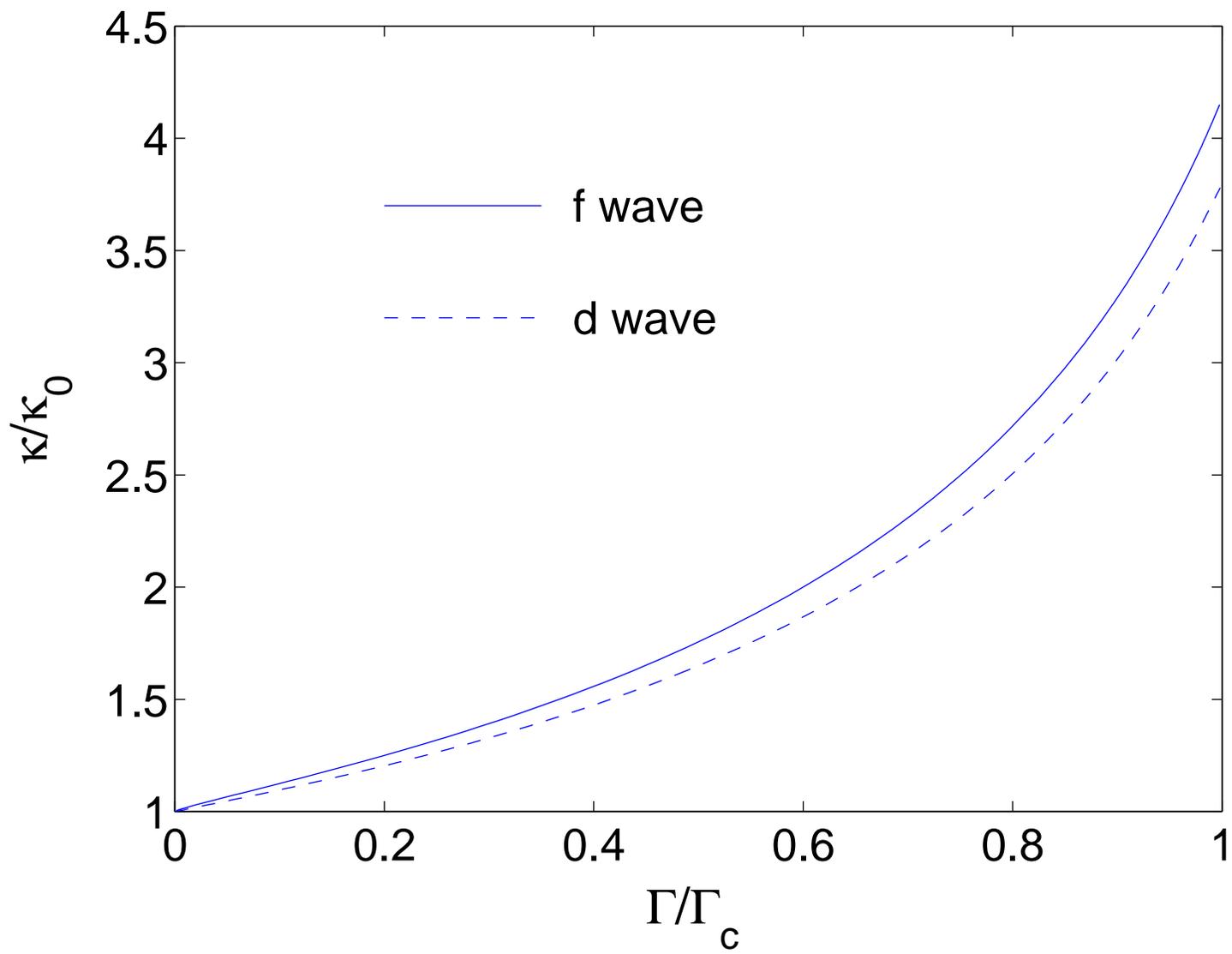
\end{document}